\documentclass[conference]{IEEEtran}
\IEEEoverridecommandlockouts
\usepackage{cite}
\usepackage{amsmath,amssymb,amsfonts}
\usepackage{algorithmic}
\usepackage{graphicx}
\usepackage{textcomp}
\usepackage{xcolor}
\usepackage{multirow}
\def\BibTeX{{\rm B\kern-.05em{\sc i\kern-.025em b}\kern-.08em
    T\kern-.1667em\lower.7ex\hbox{E}\kern-.125emX}}
\begin{document}

\title{Deep Local and Global Spatiotemporal Feature Aggregation for Blind Video Quality Assessment
}

\author{Wei Zhou,~\IEEEmembership{Student Member,~IEEE}, and Zhibo Chen,~\IEEEmembership{Senior Member,~IEEE}
\thanks{W. Zhou and Z. Chen are with the CAS Key Laboratory of Technology in Geo-Spatial Information Processing and Application System, University of Science and Technology of China, Hefei 230027, China, (weichou@mail.ustc.edu.cn; chenzhibo@ustc.edu.cn).}
\thanks{This work was supported in part by NSFC under Grant U1908209, 61632001 and the National Key Research and Development Program of China 2018AAA0101400.}}

\maketitle

\begin{abstract}
In recent years, deep learning has achieved promising success for multimedia quality assessment, especially for image quality assessment (IQA). However, since there exist more complex temporal characteristics in videos, very little work has been done on video quality assessment (VQA) by exploiting powerful deep convolutional neural networks (DCNNs). In this paper, we propose an efficient VQA method named Deep SpatioTemporal video Quality assessor (DeepSTQ) to predict the perceptual quality of various distorted videos in a no-reference manner. In the proposed DeepSTQ, we first extract local and global spatiotemporal features by pre-trained deep learning models without fine-tuning or training from scratch. The composited features consider distorted video frames as well as frame difference maps from both global and local views. Then, the feature aggregation is conducted by the regression model to predict the perceptual video quality. Finally, experimental results demonstrate that our proposed DeepSTQ outperforms state-of-the-art quality assessment algorithms.
\end{abstract}


\section{Introduction}
With the rapid growth of visual multimedia applications, evaluating the perceptual quality of multimedia data has attracted increasing attention in both academia and industry \cite{chen2018blind}. Compared with image quality assessment (IQA), how to assess video quality is more challenging due to the additional temporal dimension. Moreover, the viewed videos often consist of different visually annoying distortions that are introduced during the processing chain of digital videos including capture, compression, transmission, reconstruction, etc. Therefore, the construction of accurate video quality assessment (VQA) methods is significant for optimizing existing video services.

In general, the most reliable VQA method is to design subjective tests \cite{zhou20163d}. During the subjective tests, human subjects are asked to watch videos and then provide the quality ratings for these videos. However, the subjective quality assessment is usually labor-intensive and time-consuming, which is not applicable in practical application scenarios. Thus, it is desirable to develop efficient objective VQA algorithms to predict the perceptual quality of videos automatically.

\begin{figure}[t]
  \centerline{\includegraphics[width=8cm]{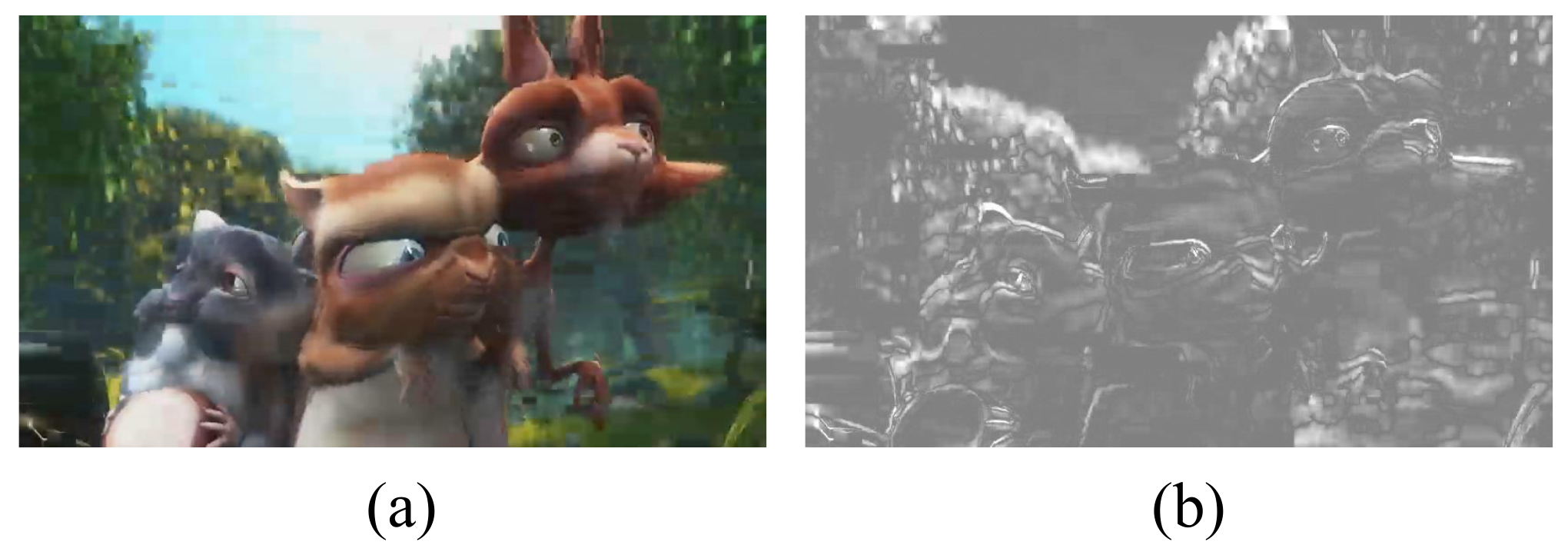}}
  \caption{Examples of distorted video frame and frame difference map from the CSIQ video quality database \cite{vu2014vis3}. (a) Current video frame, (b) Frame difference map between the current frame and the previous frame.}
  \centering
\label{fig:figure1}
\end{figure}

\begin{figure*}[t]
  \centerline{\includegraphics[width=15cm]{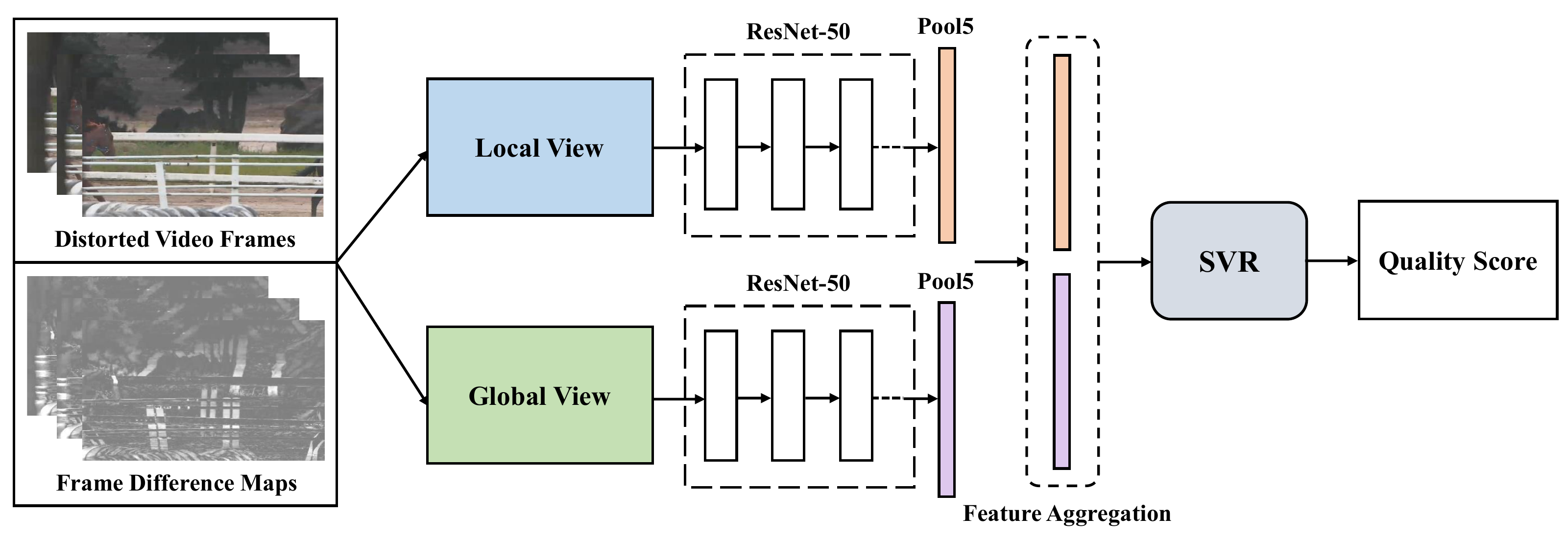}}
  \caption{The framework of our proposed DeepSTQ method.}
  \centering
\label{fig:figure2}
\end{figure*}

Depending on the availability of originally non-distorted videos, objective VQA methods can be generally classified into three categories, namely full-reference (FR) VQA, reduced-reference (RR) VQA, and no-reference/blind (NR) VQA models. The FR VQA methods always need the full information of pristine reference videos to perform quality assessment. Trational FR IQA metrics, such as peak signal-to-noise ratio (PSNR), structural similarity (SSIM) index \cite{wang2004image}, multiscale SSIM (MS-SSIM) \cite{wang2003multiscale}, feature similarity (FSIM) index \cite{zhang2011fsim}, are designed for assessing the perceptual quality of images rather than more complex videos. In the literature, several FR VQA algorithms have been proposed, which include the motion-based video integrity evaluation (MOVIE) index \cite{seshadrinathan2010motion}, the spatiotemporal most-apparent-distortion (ST-MAD) model \cite{vu2011spatiotemporal}, the algorithm for video quality assessment via analysis of spatial and spatiotemporal slices (ViS3) \cite{vu2014vis3}, the just noticeable difference-based video quality (JVQ) index \cite{loh2018just}, etc. The RR VQA methods require only part of original content. Typical RR VQA approaches include the video quality model (VQM) \cite{pinson2004new}, the spatiotemporal RR entropic differences (STRRED) algorithm \cite{soundararajan2013video}, and so on. Contrary to FR and RR VQA models, the NR VQA methods assess the perceptual video quality without any information of original videos. Consequently, the NR VQA task is more attractive since the pristine reference content is not always accessible in practical applications.

Recently, several studies have been carried out on NR VQA methods. In \cite{xu2014no}, the codebook representation for no-reference image assessment (CORNIA) \cite{ye2012unsupervised} is directly extended to NR video quality evaluation, where the V-CORNIA is proposed by frame-level unsupervised feature learning and hysteresis temporal pooling. Moreover, a spatiotemporal quality assessment model of natural video scenes in the discrete cosine transform (DCT) domain i.e. video blind image integrity notator using DCT statistics (V-BLIINDS) \cite{saad2014blind} is presented, which is derived from the image-based index called BLIINDS \cite{saad2010dct}. In \cite{mittal2016completely}, the video intrinsic integrity and distortion evaluation oracle (VIIDEO) is proposed by employing a variety of space-time statistical regularities and probing into intrinsic properties of space-time band pass video correlations. Additionally, the deep blind video quality assessment (DeepBVQA) method \cite{ahn2018deep} is proposed based on spatial features extracted from pre-trained deep learning models and hand-crafted temporal features. Nevertheless, none of the above-mentioned algorithms have utilized both the local and global spatiotemporal features by pre-trained deep learning models.

Additionally, deep convolutional neural networks (DCNNs) have shown enormous advances for many image processing and computer vision tasks. Since existing off-the-shelf DCNNs are trained on large-scale image databases with diverse image content such as ImageNet \cite{deng2009imagenet}, they could have the remarkable ability to extract discriminative image feature representation for quality assessment. Different from other quality assessment algorithms by fine-tuning or training models from scratch \cite{zhou2018stereoscopic}, exploiting the generic image feature representation extracted from pre-trained deep learning models is simple and efficient. Therefore, in this paper, we propose a blind perceptual video quality evaluation method based on local and global spatiotemporal features from distorted video frames and frame difference maps, which are extracted from off-the-shelf DCNNs.

Fig. \ref{fig:figure1} shows the examples of RGB distorted video frame and frame difference map from the CSIQ video quality database \cite{vu2014vis3}. Here, the frame difference map means the difference between current video frame and previous video frame. It should be noted that we add 128 to each pixel value in the frame difference map for better visualization. Besides, we can see that the distorted video frame represents spatial texture characteristic, while the frame difference map reveals motion information in a sense.

We describe the details of our proposed method in Section II. Section III presents the experimental results. We conclude the paper in Section IV.

\section{Proposed Method}
\begin{figure*}[t]
  \centerline{\includegraphics[width=15cm]{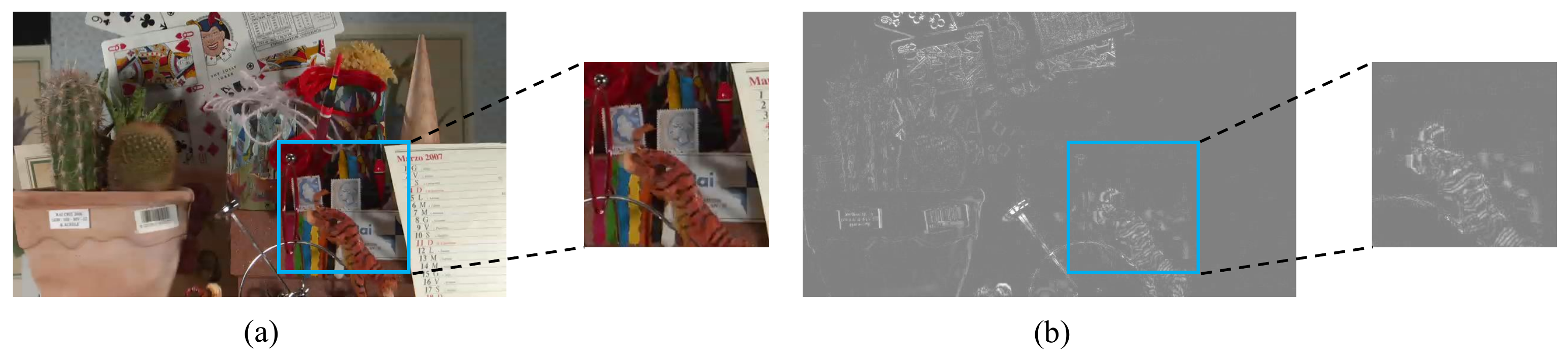}}
  \caption{Global and local views of distorted video frame and frame difference map. (a) Current video frame and the corresponding video patch, (b) Frame difference map between current frame and previous frame as well as the corresponding frame difference patch.}
  \centering
\label{fig:figure3}
\end{figure*}

In this section, we provide a detailed description of our proposed no-reference perceptual video quality evaluation method named Deep SpatioTemporal video Quality assessor (DeepSTQ). The framework of the proposed DeepSTQ method is shown in Fig. \ref{fig:figure2}. First, we generate distorted video frames and frame difference maps from both local and global views. Second, we utilize pre-trained deep learning models to extract multi-view spatiotemporal features. Finally, by aggregating the extracted features, we are able to regress them onto perceptual quality scores.

\subsection{Local and Global Spatiotemporal Representation}
Considering that a distorted video is composed of many distorted video frames, we first generate different video frames to represent the spatial texture information of the entire video. Then, the frame difference map reflecting motion information is computed based on the gray-scale distorted maps as follows:
\begin{equation}\label{1}
{{F}_{i+1}}=|{{D}_{i+1}}-{{D}_{i}}|,
\end{equation}
where ${{D}_{i+1}}$ and ${{D}_{i}}$ denote the current frame and previous frame.

Since the global and local views of an image are both important for quality assessment, we thus take the whole image as global view and the image patch as local view. Specifically, as shown in Fig. \ref{fig:figure3}, we give an example of global and local views for distorted video frame and frame difference map. The global view reflects the entire information of the image, while the local view reveals the local description of distortions.

\subsection{Deep Feature Extraction and Aggregation}
In our designed model, we employ the powerful residual network, i.e. ResNet-50 \cite{he2016deep} for deep feature extraction. The pool5 layer of this network is taken as the feature representation, which has 2048 dimensions. Moreover, since the input size of ResNet-50 is 224 $\times$ 224, we resize the distorted video frames and frame difference maps to the fixed size and then feed them into the network for global view. Besides, as for local view, we sample patches from distorted video frames and frame difference maps with the same image size and stride equaling to 112. Therefore, there exist 112 pixels overlapping between the neighboring patches, which aims to compensate for the continuity of partial distortion areas.

After the spatiotemporal deep feature extraction from both global and local views, we then separately average the feature part of each specific distorted video. Finally, the well-known regression model, i.e. support vector regression (SVR) is applied to the aggregated feature and predict the perceptual quality score.

\section{Experiments}
The experiments are conducted on the CSIQ video quality database \cite{vu2014vis3} which consists of 12 pristine reference videos and 216 distorted videos. The distorted videos cover 6 distortion types including H.264 compression, HEVC compression, motion JPEG compression, wavelet-based compression using the snow codec, additive white noise, and H.264 videos with packet loss rate subjected to simulate wireless transmission loss. Each video in the database is in the YUV420 format with the resolution of 832 $\times$ 480 and the duration of 10 seconds. The video frame rate ranges from 24 fps to 60 fps. Subjective quality score is provided for each video as the difference mean opinion score (DMOS).

For the performance evaluation of VQA algorithms, we adopt Spearman rank-order correlation coefficient (SROCC) and Pearson linear correlation coefficient (PLCC). The SROCC aims to measure the prediction monotonicity, while the PLCC is to measure the prediction accuracy. Here, the higher SROCC and PLCC indicate better performance in terms of correlation with human subjective opinions. In principle, it should be noted that before calculating the PLCC values of objective VQA algorithms, a nonlinear logistic fitting function is applied to map the predicted quality scores to the same scales of subjective quality scores. Here, we utilize a five-parameter logistic function to the predicted quality scores for a better fit to the subjective ratings as follows:
\begin{equation}\label{2}
Q(x)=\frac{{{\beta }_{1}}-{{\beta }_{2}}}{1+{{e}^{\frac{x-{{\beta }_{3}}}{{{\beta }_{4}}}}}}+{{\beta }_{2}},
\end{equation}
where ${{\beta }_{1}}$ to ${{\beta }_{4}}$ are four free parameters to be determined during the curve fitting process. Moreover, $x$ denotes the raw objective score and $Q(x)$ is the mapped score after the nonlinear fitting process.

\renewcommand\arraystretch{0.8}
\begin{table}[t]
\centering
\caption{SROCC and PLCC performance comparison on the CSIQ video quality database \cite{vu2014vis3}.}
\label{table1}
\begin{tabular}{|c|c|c|c|}
\hline
Types & Methods & SROCC & PLCC \\
\hline
\multirow{8}{*}{FR}   & PSNR    & 0.5461 & 0.5339 \\
                      & SSIM    & 0.6946 & 0.7093 \\
                      & MS-SSIM & 0.7530 & 0.6665 \\
                      & FSIM    & 0.7392 & 0.7514 \\
                      & MOVIE   & 0.8060 & 0.7880 \\
                      & ST-MAD  & 0.7355 & 0.7237 \\
                      & ViS3    & 0.8410 & 0.8300 \\
                      & JVQ     & 0.6840 & 0.7005 \\
\hline
\multirow{2}{*}{RR}   & VQM    & 0.7890 & 0.7690 \\
                      & STRRED & 0.8129 & 0.7894 \\
\hline
\multirow{4}{*}{NR}   & V-CORNIA  & 0.8216 & 0.8315 \\
                      & V-BLIINDS & 0.8069 & 0.8228 \\
                      & VIIDEO    & 0.6498 & 0.6704 \\
                      & DeepBVQA  & 0.8472 & 0.8532 \\
                      & \textbf{Proposed DeepSTQ} & \textbf{0.8533} & \textbf{0.8578} \\
\hline
\end{tabular}
\end{table}

\begin{figure}[t]
  \centerline{\includegraphics[width=7cm]{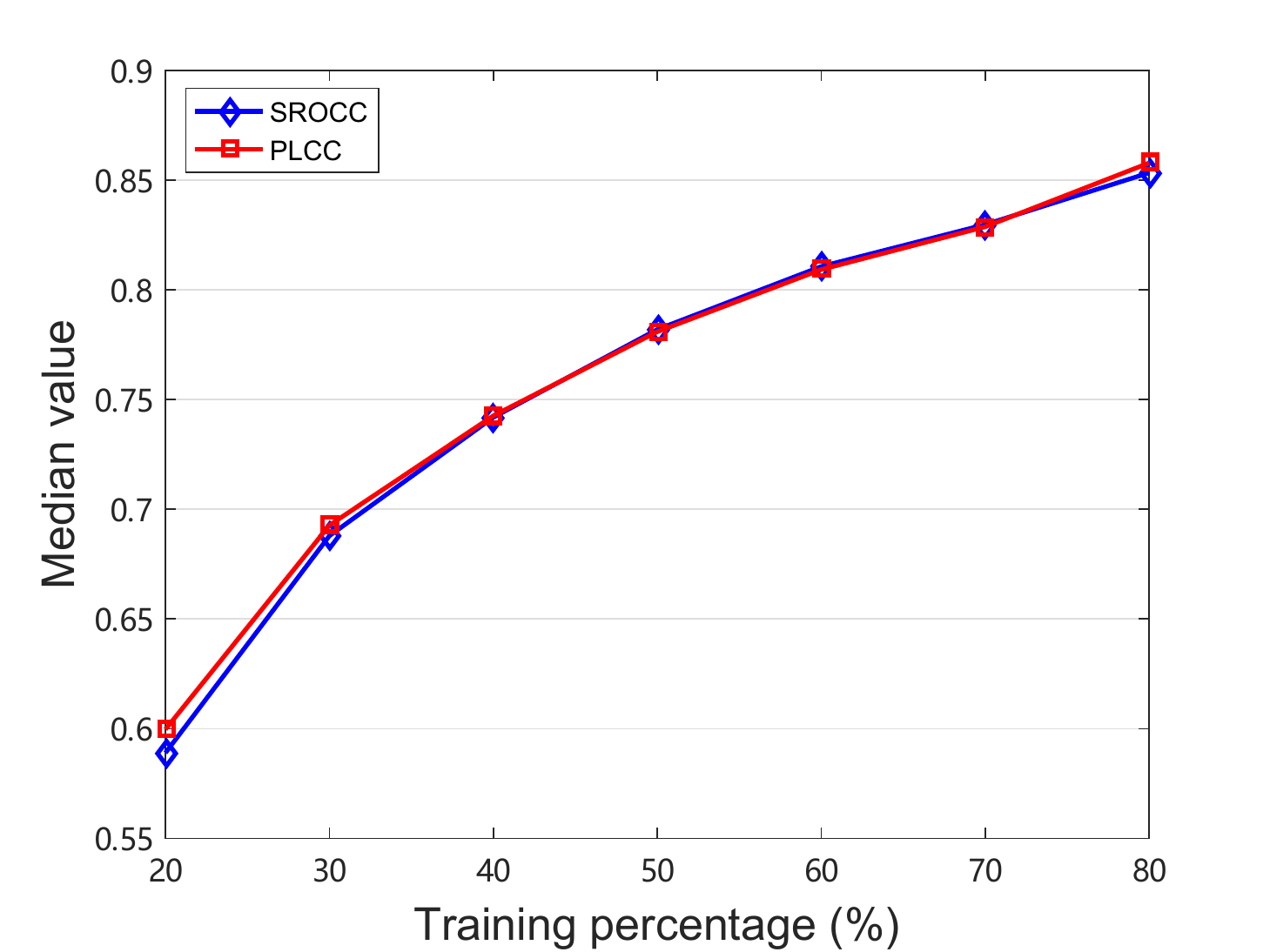}}
  \caption{Median performance in regard to training percentage over 1000 iterations on the CSIQ video quality database \cite{vu2014vis3}.}
  \centering
\label{fig:figure4}
\end{figure}

We compare the performance of our proposed DeepSTQ method with state-of-the-art VQA algorithms on the CSIQ video quality database \cite{vu2014vis3}. The 80\% of the database are randomly chosen for training and the remaining 20\% are used for testing where no overlap exists between the training and test sets. We repeat 1000 iterations of cross correlation, and then give the median SROCC and PLCC as the final results. As shown in Table 1, the proposed method is compared with 14 state-of-the-art quality assessment metrics, including PSNR, SSIM \cite{wang2004image}, MS-SSIM \cite{wang2003multiscale}, FSIM \cite{zhang2011fsim}, MOVIE \cite{seshadrinathan2010motion}, ST-MAD \cite{vu2011spatiotemporal}, ViS3 \cite{vu2014vis3}, JVQ \cite{loh2018just}, VQM \cite{pinson2004new}, STRRED \cite{soundararajan2013video}, V-CORNIA \cite{xu2014no}, V-BLIINDS \cite{saad2014blind}, VIIDEO \cite{mittal2016completely}, and DeepBVQA \cite{ahn2018deep}. The DeepBVQA is a deep learning based approach to predict the quality of distorted videos without reference. It should be noted that PSNR, SSIM, MS-SSIM and FSIM are FR IQA metrics where we compute for each video frames and then average to derive the final performance. From Table 1, we can see that our proposed DeepSTQ outperforms state-of-the-art algorithms, which demonstrates the effectiveness of the proposed method. Fig. 4 shows SROCC and PLCC performance variation with respect to training percentage on the CSIQ video quality database \cite{vu2014vis3}. We find that a large number of training data bring about performance increase.

Furthermore, in order to investigate the specific contribution of each technical part for our proposed DeepSTQ, we present the ablation study of the proposed method on the CSIQ video quality database \cite{vu2014vis3}, as shown in Table 2. As for each method, we extract features from the same pool5 layer of the pre-trained ResNet-50 model, and then carry out perceptual quality regression. First, we observe that the extracted features of frame difference maps perform better than that of distorted video frames. One possible explanation may be that motion information has more influence than spatial texture on perceptual video quality. Besides, using the combined features from distorted video frames and frame difference maps outperforms using either the two types of features alone, which verifies the importance of composited spatiotemporal features. Second, contrary to the previous observation, the performance of extracted features from distorted video patches is superior to that of frame difference patches. This is because some of the cropped patches for frame difference maps lost abundant texture details, which are visualized as a whole gray image. Likewise, using the combined features from distorted video patches and frame difference patches also shows advantage than using either the two types of features alone. Finally, the result on local view metrics is better than that on global view metrics with the same setting. This demonstrates that the local discriminative features seem to be more important for quality assessment. The proposed DeepSTQ, which uses composited spatiotemporal features both from global and local views, achieves higher SROCC and PLCC performance values.

\renewcommand\arraystretch{0.8}
\begin{table}[t]
\centering
\caption{Ablation study of the proposed method on the CSIQ video quality database \cite{vu2014vis3}.}
\label{table2}
\begin{tabular}{|c|c|c|}
\hline
Methods & SROCC & PLCC \\
\hline
Distorted video frames & 0.7915 & 0.8031 \\
Frame difference maps  & 0.8113 & 0.8177 \\
Distorted video frames + Frame difference maps & 0.8224 & 0.8307 \\
\hline
Distorted video patches  & 0.8232 & 0.8298 \\
Frame difference patches & 0.8175 & 0.8263 \\
Distorted video patches + Frame difference patches & 0.8503 & 0.8551 \\
\hline
\textbf{Proposed DeepSTQ} & \textbf{0.8533} & \textbf{0.8578} \\
\hline
\end{tabular}
\end{table}

\section{Conclusion}
In this paper, we present an efficient blind video quality evaluation method based on deep learning models. Specifically, we first exploit pre-trained off-the-shelf DCNNs to generate the discriminative features of distorted video frames as well as frame difference maps from both local and global views. The proposed DeepSTQ considers the spatiotemporal characteristics of various distorted videos. We then use the regression model to aggregate features and assess perceptual video quality. Experimental results demonstrate that our proposed DeepSTQ achieves superior performance compared to other state-of-the-art VQA algorithms. In the future, we plan to study 3D video quality evaluation based on deep learning.

\bibliographystyle{IEEEbib}
\bibliography{references}

\begin{thebibliography}{10}

\bibitem{chen2018blind}
Zhibo Chen, Wei Zhou, and Weiping Li,
\newblock ``Blind stereoscopic video quality assessment: From depth perception
  to overall experience,''
\newblock {\em IEEE Transactions on Image Processing}, vol. 27, no. 2, pp.
  721--734, 2018.

\bibitem{zhou20163d}
Wei Zhou, Ning Liao, Zhibo Chen, and Weiping Li,
\newblock ``{3D-HEVC} visual quality assessment: Database and bitstream
  model,''
\newblock in {\em Quality of Multimedia Experience (QoMEX), 2016 Eighth
  International Conference on}. IEEE, 2016, pp. 1--6.

\bibitem{vu2014vis3}
Phong~V Vu and Damon~M Chandler,
\newblock ``{ViS3}: an algorithm for video quality assessment via analysis of
  spatial and spatiotemporal slices,''
\newblock {\em Journal of Electronic Imaging}, vol. 23, no. 1, pp. 013016,
  2014.

\bibitem{wang2004image}
Zhou Wang, Alan~C Bovik, Hamid~R Sheikh, and Eero~P Simoncelli,
\newblock ``Image quality assessment: from error visibility to structural
  similarity,''
\newblock {\em IEEE Transactions on image processing}, vol. 13, no. 4, pp.
  600--612, 2004.

\bibitem{wang2003multiscale}
Zhou Wang, Eero~P Simoncelli, and Alan~C Bovik,
\newblock ``Multiscale structural similarity for image quality assessment,''
\newblock in {\em The Thrity-Seventh Asilomar Conference on Signals, Systems \&
  Computers, 2003}. IEEE, 2003, vol.~2, pp. 1398--1402.

\bibitem{zhang2011fsim}
Lin Zhang, Lei Zhang, Xuanqin Mou, and David Zhang,
\newblock ``{FSIM}: A feature similarity index for image quality assessment,''
\newblock {\em IEEE Transactions on Image Processing}, vol. 20, no. 8, pp.
  2378--2386, 2011.

\bibitem{seshadrinathan2010motion}
Kalpana Seshadrinathan and Alan~Conrad Bovik,
\newblock ``Motion tuned spatio-temporal quality assessment of natural
  videos,''
\newblock {\em IEEE Transactions on image processing}, vol. 19, no. 2, pp.
  335--350, 2010.

\bibitem{vu2011spatiotemporal}
Phong~V Vu, Cuong~T Vu, and Damon~M Chandler,
\newblock ``A spatiotemporal most-apparent-distortion model for video quality
  assessment,''
\newblock in {\em Image Processing (ICIP), 2011 18th IEEE International
  Conference on}. IEEE, 2011, pp. 2505--2508.

\bibitem{loh2018just}
Woei-Tan Loh and David Boon~Liang Bong,
\newblock ``A just noticeable difference-based video quality assessment method
  with low computational complexity,''
\newblock {\em Sensing and Imaging}, vol. 19, no. 1, pp. 33, 2018.

\bibitem{pinson2004new}
Margaret~H Pinson and Stephen Wolf,
\newblock ``A new standardized method for objectively measuring video
  quality,''
\newblock {\em IEEE Transactions on broadcasting}, vol. 50, no. 3, pp.
  312--322, 2004.

\bibitem{soundararajan2013video}
Rajiv Soundararajan and Alan~C Bovik,
\newblock ``Video quality assessment by reduced reference spatio-temporal
  entropic differencing,''
\newblock {\em IEEE Transactions on Circuits and Systems for Video Technology},
  vol. 23, no. 4, pp. 684--694, 2013.

\bibitem{xu2014no}
Jingtao Xu, Peng Ye, Yong Liu, and David Doermann,
\newblock ``No-reference video quality assessment via feature learning,''
\newblock in {\em Image Processing (ICIP), 2014 IEEE International Conference
  on}. IEEE, 2014, pp. 491--495.

\bibitem{ye2012unsupervised}
Peng Ye, Jayant Kumar, Le~Kang, and David Doermann,
\newblock ``Unsupervised feature learning framework for no-reference image
  quality assessment,''
\newblock in {\em 2012 IEEE conference on computer vision and pattern
  recognition}. IEEE, 2012, pp. 1098--1105.

\bibitem{saad2014blind}
Michele~A Saad, Alan~C Bovik, and Christophe Charrier,
\newblock ``Blind prediction of natural video quality,''
\newblock {\em IEEE Transactions on Image Processing}, vol. 23, no. 3, pp.
  1352--1365, 2014.

\bibitem{saad2010dct}
Michele~A Saad, Alan~C Bovik, and Christophe Charrier,
\newblock ``A {DCT} statistics-based blind image quality index,''
\newblock {\em IEEE Signal Processing Letters}, vol. 17, no. 6, pp. 583--586,
  2010.

\bibitem{mittal2016completely}
Anish Mittal, Michele~A Saad, and Alan~C Bovik,
\newblock ``A completely blind video integrity oracle,''
\newblock {\em IEEE Transactions on Image Processing}, vol. 25, no. 1, pp.
  289--300, 2016.

\bibitem{ahn2018deep}
Sewoong Ahn and Sanghoon Lee,
\newblock ``Deep blind video quality assessment based on temporal human
  perception,''
\newblock in {\em 2018 25th IEEE International Conference on Image Processing
  (ICIP)}. IEEE, 2018, pp. 619--623.

\bibitem{deng2009imagenet}
Jia Deng, Wei Dong, Richard Socher, Li-Jia Li, Kai Li, and Li~Fei-Fei,
\newblock ``Imagenet: A large-scale hierarchical image database,''
\newblock in {\em Computer Vision and Pattern Recognition, 2009. CVPR 2009.
  IEEE Conference on}. Ieee, 2009, pp. 248--255.

\bibitem{zhou2018stereoscopic}
Wei Zhou, Zhibo Chen, and Weiping Li,
\newblock ``Stereoscopic video quality prediction based on end-to-end dual
  stream deep neural networks,''
\newblock in {\em Pacific Rim Conference on Multimedia}. Springer, 2018, pp.
  482--492.

\bibitem{he2016deep}
Kaiming He, Xiangyu Zhang, Shaoqing Ren, and Jian Sun,
\newblock ``Deep residual learning for image recognition,''
\newblock in {\em Proceedings of the IEEE conference on computer vision and
  pattern recognition}, 2016, pp. 770--778.

\end{thebibliography}

\end{document}